\documentclass[%
 reprint,
 amsmath,amssymb,
 aps,
]{revtex4-2}

\usepackage{graphicx}% Include figure files
\usepackage{dcolumn}% Align table columns on decimal point
\usepackage{bm}% bold math
\usepackage{xcolor}
\usepackage{gensymb}
\definecolor{SIgreen}{HTML}{50C878}

\definecolor{katrijnpink}{HTML}{FF007F}

\begin{document}

\preprint{APS/123-QED}
\title{AC magnetometry in the strong drive regime with NV centers in diamond}% Force line breaks with \\

\author{Katrijn Everaert$^{1,2}$, Saipriya Satyajit$^{1,2}$, Jiashen Tang$^{1,2}$, Zechuan Yin$^{1,3}$, Xiechen Zheng$^{1,3}$\\Jner Tzern Oon$^{1,2}$, Connor A. Hart$^{1,3}$, John W. Blanchard$^{1,3}$, Ronald L. Walsworth$^{1,2,3}$}
\affiliation{$^{1}$Quantum Technology Center, University of Maryland, College Park, Maryland 20742, USA
}%

\affiliation{$^{2}$Department of Physics, University of Maryland, College Park, Maryland 20742, USA
}%
\affiliation{$^{3}$Department of Electrical Engineering and Computer Science, University of Maryland, College Park,
Maryland 20742, USA
}%

%\date{}% It is always \today, today,

\begin{abstract}
Magnetic response measurements in the presence of AC drive fields provide critical insight into the properties of magnetic and conductive materials, such as phase transitions in two-dimensional van der Waals magnets, the heating efficiency of magnetic nanoparticles in biological environments, and the integrity of metals in eddy current testing. Nitrogen-vacancy (NV) centers in diamond are a commonly-used platform for such studies, due to their high spatial resolution and sensitivity, but are typically limited to weak-drive conditions, i.e., AC drive fields $B_D$ well below the NV microwave (MW) pulse Rabi strength $\Omega$. Once $B_D$ grows comparable to or larger than $\Omega$, the induced MW pulse detuning suppresses NV sensitivity to the out-of-phase magnetic response, which encodes dissipation and conductivity in materials of interest. Here, we introduce a phase modulation protocol that cancels MW pulse detuning to leading order, and extends NV AC magnetometry into the strong drive field regime. The protocol, termed SIPHT (Signal Isolation through PHase Tuning), is experimentally demonstrated using an NV ensemble. By directly comparing SIPHT to the conventional Hahn echo AC sensing protocol, we quantify the preservation of NV magnetometry contrast for an out-of-phase signal. We further showcase SIPHT by detecting eddy current-induced magnetic fields from Cu, Al, and Ti samples, with the measured response field phase delays reflecting their distinct conductivities. SIPHT extends NV AC magnetometry to regimes inaccessible to standard dynamical decoupling measurement protocols, unlocking novel utility, e.g., in the study of magnetic hyperthermia and nondestructive testing of conductors. 
\end{abstract}

\maketitle

%\tableofcontents

\section{Introduction}

A broad range of applications rely on the ability to measure the response of materials to alternating magnetic fields. Understanding the phase transitions and their related properties in 2D van der Waals (vdW) magnets is facilitated by magnetic AC susceptibility measurements \cite{Meseguer-Snchez2021,Zhang2021,Tschudin2024}. Assessment of the integrity of conductive materials can be achieved in a non-invasive fashion via eddy current detection \cite{Garca-Martn2011,Wickenbrock2016,Chatzidrosos2019}. In the biomedical field, magnetic hyperthermia is a promising cancer therapy that involves exposing magnetic nanoparticles (MNPs) to strong alternating magnetic fields in the mT range, causing localized heating to selectively destroy cancer tissue \cite{Rosensweig2002,Rodrigues2020,Rubia-Rodrguez2021}. Characterizing and optimizing the magnetic response of MNPs in biological environments is essential to ensure that this heat can be generated efficiently, precisely, and safely \cite{Cabrera2018,Gaviln2021,Sharifabad2024}. A key aspect of the magnetic response signal is the out-of-phase component, which encodes critical information about energy dissipation and magnetic relaxation. For example, in magnetic hyperthermia, AC hysteresis losses determine the heating efficiency of MNPs \cite{Carrey2011,Dennis2013,Everaert2024}. Similarly, in eddy current testing, the phase delay of the response field reveals variations in electrical conductivity and highlights structural anomalies.\\

Nitrogen-vacancy (NV) centers in diamond are widely used to sense and image magnetic signals from a variety of materials and systems across the physical and life sciences, offering high spatial resolution and sensitivity, as well as compatibility with diverse samples  \cite{Taylor2008, Rovny2024}, which makes them highly attractive for the applications discussed above. In the context of MNPs, NV-based techniques enable sub-micrometer magnetic imaging \cite{LeSage2013, Glenn2015}, allowing the characterization of static and dynamic properties of single particles \cite{ SchaferNolte2014, Richards2025}. This capability is critical for studying how biological environments influence the magnetization dynamics of MNPs at the cellular-level - information that is inaccessible using conventional sensing methods such as SQUIDs or atom-based magnetometers. Extending NV sensing capabilities to include the response of MNPs to strong AC drive fields could provide valuable insights into interparticle interactions, aggregation phenomena, and how local variations in viscosity and temperature affect the hyperthermia performance of MNPs  \cite{Cabrera2018, Sharifabad2024}.\\

NV AC sensing typically leverages dynamical decoupling (DD) pulse sequences using applied microwave (MW) pulses to coherently manipulate NV electronic spins, enabling magnetometry in the $\sim$ 1 kHz - 10 MHz range \cite{Levine2019, Barry2020}. To date, however, DD-based NV AC measurements have been limited to small AC drive fields \cite{Zhang2021}, particularly when probing the out-of-phase magnetic response signal from materials of interest, for which the MW pulses used for NV spin control are applied at the drive field antinodes. Strong drive fields typically shift the NV spin resonance, resulting in MW detuning that degrades the sensitivity of conventional DD-based NV AC magnetometry.\\
\begin{figure*}
\centering
  \includegraphics[width=6.5in]{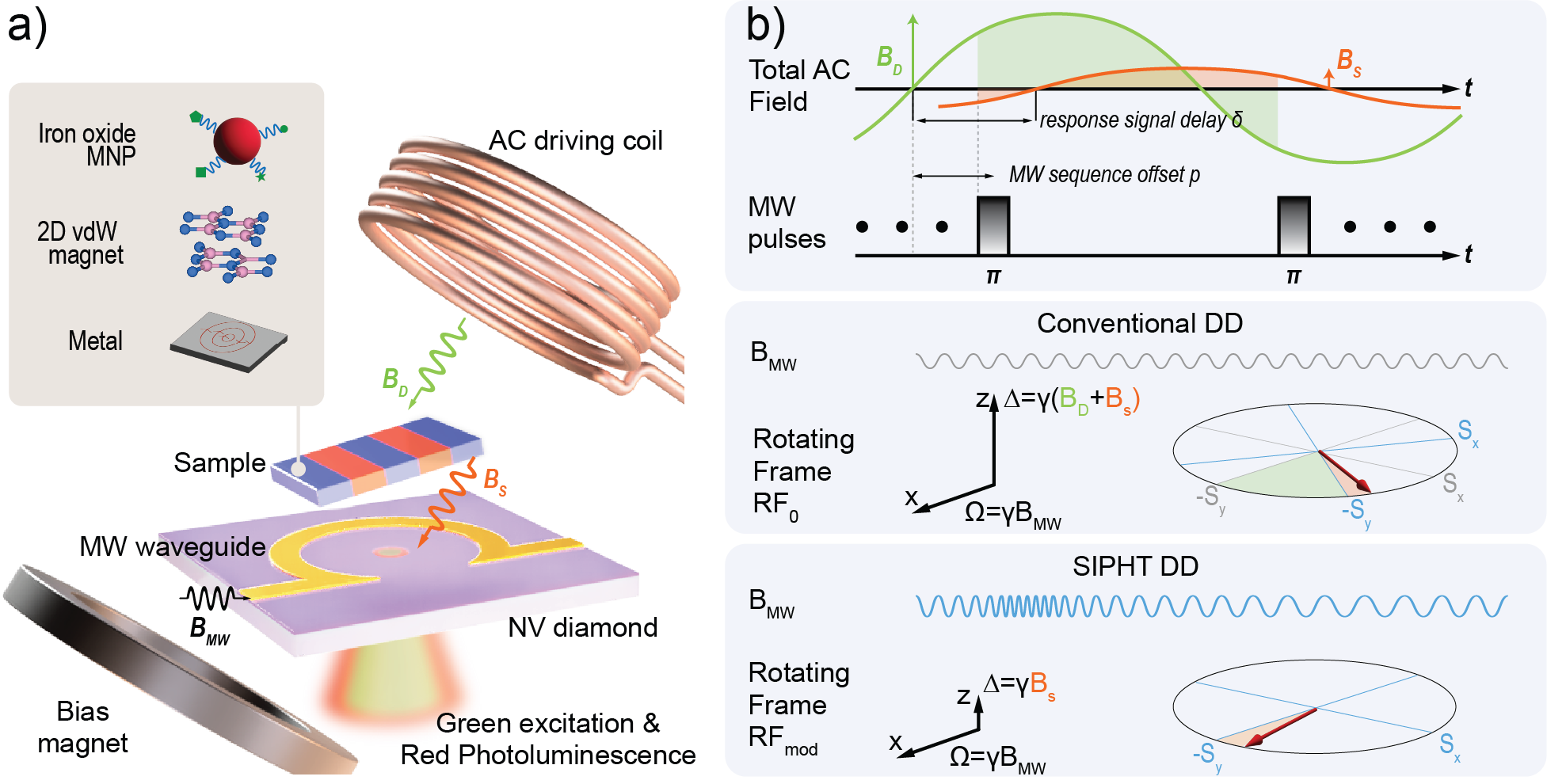}
  \caption{SIPHT dynamical decoupling (DD) for AC magnetometry of magnetic response signals in the strong drive regime. a) Experimental schematic for an AC response measurement with NV centers in diamond. A strong AC drive field ($B_D$) generated by a coil induces an AC response field ($B_S$) from a sample (with examples shown in the shaded inset), sensed by an NV ensemble. MW pulses for NV spin control during DD-based AC magnetometry measurements are generated by an $\Omega$-shaped waveguide. The static bias magnetic field and AC drive field are aligned to the probed NV axis. b) NV electronic spin phase accumulation in conventional and SIPHT DD. In conventional DD, NV spins precess around $\Delta/\gamma=(B_D+B_S)$ in rotating frame RF$_0$ and accumulate phase due to both $B_D$ and $B_S$, highlighted by the green and orange areas in the Bloch plane. In SIPHT DD, rotating frame RF$_{\text{mod}}$ has a tunable phase modulation, which can be chosen to match the NV phase accumulation due to $B_D$ in RF$_0$. When this condition is satisfied, the SIPHT DD sequence effectively isolates the signal field $B_S$ by rejecting all contributions from $B_D$ to NV phase accumulation.}
  \label{fig:sipht}
\end{figure*}

In this work, we introduce a MW phase modulation scheme into an NV DD protocol that cancels leading-order contributions from an AC drive field, thereby allowing sensitive NV AC magnetometry in the strong drive field regime. This approach, which we term SIPHT (Signal Isolation through PHase Tuning) restores sensitivity to the out-of-phase component of an AC magnetic signal even under large drive fields. In Sec. \ref{sec:results}, we first outline the principle of SIPHT and its implementation in a Hahn Echo sequence using NV ensembles; then benchmark SIPHT performance experimentally against the conventional AC sensing protocol, validate its robustness, and finally apply it to measure response signals arising from eddy currents in metallic samples. In Sec. \ref{sec:discussion}, we discuss potential extensions of SIPHT for other NV DD sequences and broader sensing applications. Sec. \ref{sec:methods} details the experimental implementation.

\section{Results\label{sec:results}}
\subsection{SIPHT dynamic decoupling for AC response magnetometry}

We consider a material (examples in Fig. \ref{fig:sipht}a) subjected to a large sinusoidal AC magnetic drive field given by $B_{D}\cos{(2\pi f_{D} t)}$ where $B_D$ is the amplitude and $f_{D}$ the frequency. This drive field induces a small magnetic AC response from the material along the same direction, described by $B_{S}\cos{(2\pi f_{D} t-\delta)}$ where $B_S$ is the response field amplitude and $\delta$ the phase delay with respect to the drive field, as shown in Fig. \ref{fig:sipht}b. The total AC magnetic field is the sum of the drive and response fields: $B_{D}\cos{(2\pi f_{D} t)}+B_{S}\cos{(2\pi f_{D} t-\delta)}$. The objective is to determine $B_S$ and $\delta$, in particular for situations where $B_D$ is comparable or larger than the Rabi strength $\Omega$ of MW pulses used for NV spin control, e.g., in DD-based AC magnetometry. The drive and response fields considered are parallel to the NV axis used for sensing.\\

The proposed AC magnetometry protocol (SIPHT DD) efficiently isolates weak AC response signals from strong drive fields by introducing phase tuning of the MW pulses used in a DD sensing sequence. Specifically, we replace the conventional MW pulses in the DD sequence with a phase-modulated version:

\begin{equation}
    B_{\text{MW}}=\frac{\Omega}{\gamma} \cos\left((D-\gamma B_{DC})t+\gamma B_{D}'\frac{\sin(2\pi f_{D}t)}{2\pi f_{D}}\right)\label{eq:MW_SIPHT}
\end{equation}
with $\gamma \approx 2\pi \cdot28 $ GHz/T the gyromagnetic ratio of the NV electronic spin, $D \approx$ $2\pi \cdot2.87$ GHz the NV zero field spitting, $B_{DC}$ the static bias field along the NV axis of symmetry, and $B_D'$ the phase modulation amplitude. In conventional DD, $B_D'=0$, and the NV electronic spin accumulates phase in the rotating frame RF$_{0}$ due to both $B_D$ and $B_S$. When $B_D'\neq0$, the rotating frame RF$_{\text{mod}}$ has an additional time-dependent phase $B_{D}'\frac{\sin(2\pi f_{D}t)}{2\pi f_{D}}$ with respect to RF$_{0}$. The phase accumulation $\phi_{\text{NV}}$ at the end of the SIPHT DD sequence (see Supplementary Information) is given by
\begin{equation}
    \phi_{\text{NV}} = \frac{2 \gamma N_{\pi}}{\pi f_{D}}\left[(B_D-B_D')\cos(p)+B_S\cos(p-\delta)\right]\label{eq:phase_SIPHT}
\end{equation}
with $N_{\pi}$ the number of $\pi$ pulses in the DD sequence. $p$ is the phase offset of the MW pulse train with respect to the drive field nodes; and $\delta$ is the phase delay of the signal field with respect to the drive field as shown in Fig. \ref{fig:sipht}b. After the last $\pi/2$ MW pulse, the NV measurement contrast $C\propto\sin{(\phi_{\text{NV}})}$ is read out as a photoluminescence (PL) signal.\\

With precise MW pulse phase control using an AWG, the phase of RF$_{\text{mod}}$ w.r.t. RF$_0$ can be matched to the phase accumulated by the NV spin in RF$_0$ due to $B_{D}$ by assuring that $B_{D}'=B_{D}$ (details in Sec. \ref{sec:methods} and the Supplementary Information). As a result, the NV spins in RF$_{\text{mod}}$ do not accumulate net phase due to $B_D$ during the SIPHT DD sequence and are thus effectively insensitive to the drive field $B_{D}$. In the presence of a magnetically responsive material, only the response signal $B_{S}$ contributes to NV spin phase accumulation, and thus the SIPHT DD sequence is only sensitive to $B_S$, as illustrated schematically in Fig. \ref{fig:sipht}b. We highlight that SIPHT DD cancels net NV spin phase accumulation from $B_D$ via simple MW pulse phase modulation, without a physical counter-field involved, i.e. the sample under study experiences the same AC drive field as with conventional techniques. This approach allows SIPHT DD to be applied for arbitrary values of $B_D$.\\

A second key benefit of MW pulse phase modulation under SIPHT conditions (when $B_D'=B_D$) is avoidance of (unwanted) detuning of MW pulses away from NV spin resonance, as occurs in conventional DD with strong AC drive fields. This result is achieved by eliminating the effect of the drive field in the rotating frame RF$_{\text{mod}}$, thereby preserving MW resonance with the NV spin transition, ensuring effectiveness of the MW pulses for NV spin control, and enhancing NV sensitivity to the AC response field.

\subsection{Showcasing SIPHT DD}

\begin{figure}
\includegraphics[width=\linewidth]{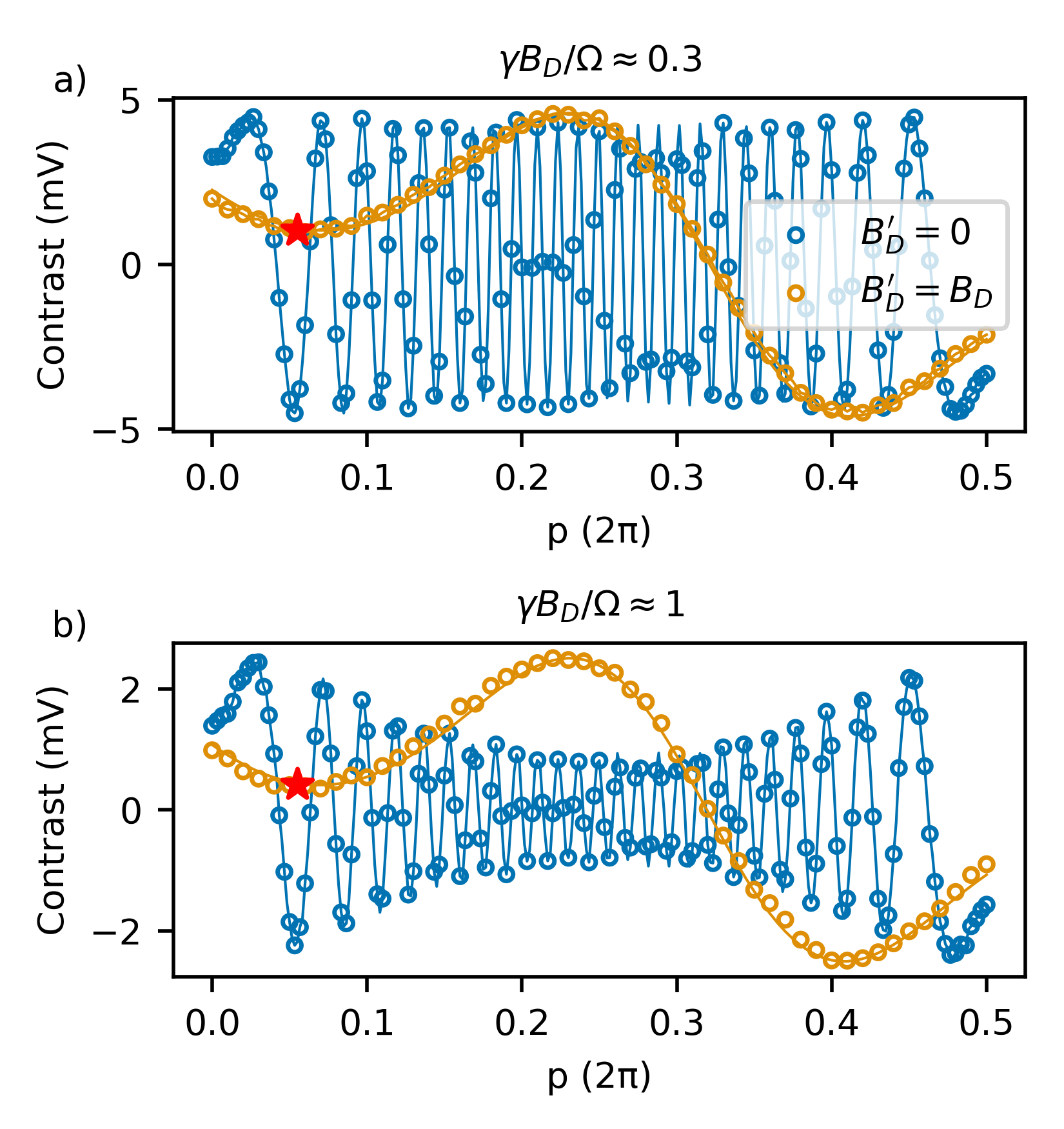}  
\caption{NV AC signal measurement contrast for a Hahn echo sequence in the presence of a coil generated field at $f_D$ = 152 kHz, consisting of a drive field $B_D \approx$ 100 $\mu$T and a response field $B_S \approx$ 4 $\mu$T at a known phase delay $\delta=0.065\cdot 2\pi$ ($\approx 23\degree$). NV AC measurements are made under SIPHT ($B_D'=B_D$) and conventional ($B_D'=0$) DD conditions for a relatively modest and large normalized drive field $\gamma B_D/\Omega$ in a) and b) respectively. For conventional DD conditions, NV spin phase accumulation due to $B_D$ induces large oscillations in the contrast as a function of the phase offset $p$ of the MW pulse train used to control the NV spins. Under SIPHT conditions, phase accumulation is solely due to $B_S$ and the phase delay $\delta$ of the response field can be read out directly from the contrast data at values of $p$ where the contrast is symmetric (red stars in a) and b)). For large normalized drive fields $\gamma B_D/\Omega$, contrast is reduced for conventional DD around values where $|B_D|$ is maximum. SIPHT preserves contrast in the strong drive field regime.}\label{fig:phase}
\end{figure}

We experimentally demonstrate the impact of a strong drive field on NV AC measurement contrast, with results shown in Fig. \ref{fig:phase}. An AC drive field of peak amplitude $B_D \approx$ 100 $\mu$T (corresponding to $\gamma B_D \approx 2\pi \cdot 2.8$ MHz) and a synthetic (coil-generated) response field of peak amplitude $B_S \approx$ 4 $\mu$T with a defined phase delay $\delta=0.065 \cdot 2\pi$ ($\approx$ 23\degree) are applied using a coil. Both the drive and response fields are at frequency $f_D =$ 152 kHz. A sweep of the phase offset $p$ of the MW pulse train, used for NV spin control, is performed under both conventional DD conditions ($B_D' = 0$) and SIPHT DD conditions ($B_D' = B_D$). A Hahn echo sequence is used for NV AC signal measurements, with two different MW pulse Rabi strengths: $\Omega = 2\pi \cdot9.6$ MHz and $\Omega = 2\pi \cdot2.8$ MHz (shown in Fig. \ref{fig:phase}a and \ref{fig:phase}b, respectively). Further details of the experimental setup and methods are described in Sec. \ref{sec:methods} and the Supplementary Information.\\

For a large drive field $B_D$ with $\gamma B_D/\Omega < 1$, NV spin phase accumulation in conventional DD is dominated by $B_D$, leading to rapid oscillations in the measurement contrast as a function of $p$ (see blue points in Fig. \ref{fig:phase}a). However, under SIPHT conditions where $B_D'=B_D$, NV spin phase accumulation is solely due to the response field; and rapid oscillations in measurement contrast as a function of $p$ are not observed (see orange points in Fig. \ref{fig:phase}a). In both cases, $B_S$ and $\delta$ can be estimated by jointly fitting them to the measurement contrast. However, under SIPHT DD conditions, $\delta$ can also be directly determined from the measurement data at values of $p$ where the contrast is symmetric, as indicated by the red star in Fig. \ref{fig:phase}a. Additionally, $B_S$ can be estimated from the number of local maxima $\eta$ in the measurement contrast over a $2\pi$ period in $p$: $0<\frac{(\eta-2)\pi^2f_D}{4\gamma}<B_S<\frac{\eta\pi^2f_D}{4\gamma}$. For the relatively modest normalized drive amplitude ($\gamma B_D/\Omega \approx 0.3$) used on the measurements shown in Fig. \ref{fig:phase}a, MW pulse detuning does not cause noticeable contrast loss for conventional DD.\\

For large values of both $B_D$ and $\gamma B_D/\Omega$, drive field-induced detuning of MW pulses at values of $p$ for which $|B_D|$ is maximum (i.e., $p \approx \pi/2$ and $p \approx 3\pi/2$) degrades the effect of MW pulses on NV spins. This degraded MW pulse fidelity is visible in Fig. \ref{fig:phase}b, where the conventional ($B_D' = 0$) DD sequence shows reduced measurement contrast near $p \approx \pi/2$. In such cases, fitting $B_S$ and $\delta$ to the measurement contrast can lead to systematic deviations from the true values. Under SIPHT conditions, there is no induced MW pulse detuning; and the NV measurement contrast is preserved for all values of $p$; fitting in this case yields accurate estimates of both $B_S$ and $\delta$. Again, $\delta$ can be directly determined from the SIPTH DD data, as indicated by the red star in Fig. \ref{fig:phase}b; and $B_S$ can be estimated from the number of local maxima over a 2$\pi$ period in $p$. The observed difference in overall measurement contrast between Figs. \ref{fig:phase}a and \ref{fig:phase}b is a result of the longer sequence duration used at lower $\Omega$ (for results in Fig \ref{fig:phase}b).

\subsection{Quantifying contrast preservation\\ with SIPHT DD}
To quantify the preservation of NV measurement contrast provided by SIPHT DD for increasing $\gamma B_{D}/\Omega$, one period $P$ of an AC magnetometry curve is recorded using MW pulses timed at $\delta=\pi/2$ by adding a second term in the MW phase modulation and sweeping its amplitude $B_S'\in[0,P=\frac{\pi f_{D}}{2\gamma}]$:

\begin{equation}
\begin{split}
    B_{\text{MW}}&=\frac{\Omega}{\gamma} \cos\Big((D-\gamma B_{DC})t+B_{D}'\frac{\sin(2\pi f_{D}t)}{2\pi f_{D}}\\
    &+B_{S}'\frac{\sin(2\pi f_{D}t-\delta)}{2\pi f_{D}}\Big)
    \end{split}
\end{equation}

\begin{figure}
\centering
  \includegraphics[width=0.50\textwidth]{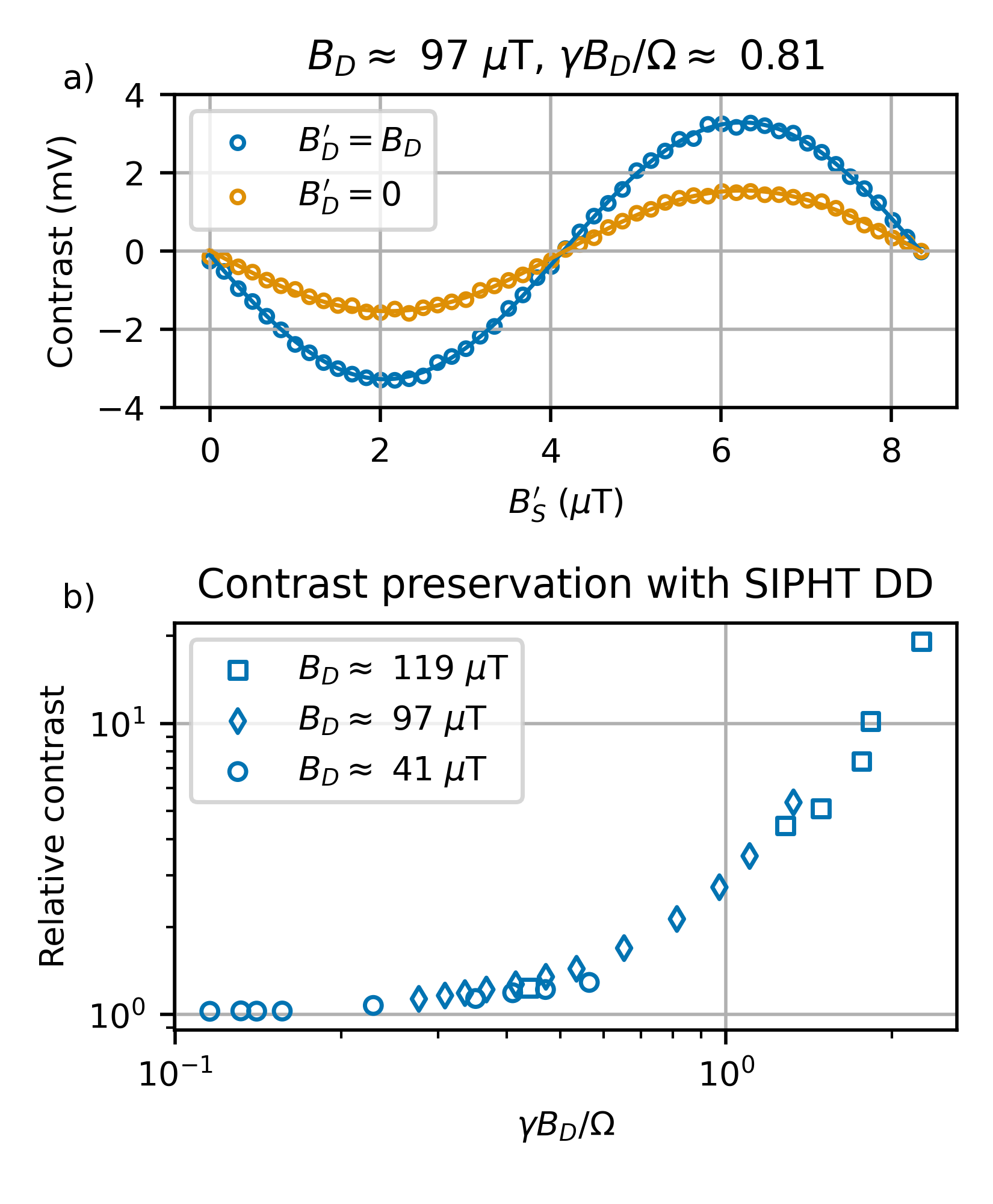}
  \caption{a) First period of the measured out-of-phase response ($p=\pi/2$) NV AC magnetometry curve generated by sweeping $B_S'$ for a drive field of $B_D\approx $ 97 $\mu$T at $f_D$ = 149 kHz, and MW pulse Rabi strength of 3.32 MHz ($\gamma B_D/\Omega \approx  0.81$).  b) Relative contrast of SIPHT Hahn echo ($B_D'=B_D$) with respect to conventional Hahn echo ($B_D'=0$) for out-of-phase response NV AC magnetometry measurements ($p=\pi/2$), as a function of normalized drive field amplitude $\gamma B_D/\Omega$. Significant contrast loss is observed for conventional DD compared to SIPHT DD at larger $\gamma B_D/\Omega$.}
  \label{fig:OOPR}
\end{figure}

With this approach, the net accumulated NV spin phase due to the response field $B_S$ is linear, producing a sinusoidal AC magnetometry curve as a function of measurement contrast. We perform such measurements under conventional ($B_D'=0$) and SIPHT ($B_D'=B_D$) DD conditions, for various values of the drive field amplitude $B_D$ and MW pulse Rabi strength $\Omega$, with $p=\delta=\pi/2$ at a drive frequency $f_D$ = 149 kHz. A Hahn echo sequence is again used for NV AC signal measurements. \\

Fig. \ref{fig:OOPR}a shows the resulting NV AC magnetometry curves for one such pair of parameters ($B_D\approx97 \mu$T, $\Omega$=$2\pi \cdot$3.32 MHz), resulting in a normalized drive field amplitude of $\gamma B_D/\Omega \approx 0.81$. For these parameters, the contrast amplitude of the SIPHT DD magnetometry curve is approximately twice that of the conventional Hahn echo. Fig. \ref{fig:OOPR}b shows the relative magnetometry amplitudes for all studied pairs of ($B_D,\Omega$) as a function of the normalized drive field amplitude $\gamma B_{D}/\Omega$ (see SI for full datasets and fits). Reduced MW pulse efficiency leads to significant observed contrast loss at larger $\gamma B_{D}/\Omega$ for conventional DD; whereas SIPHT DD suffers no such contrast degradation.\\

From these results, enhanced sensitivity to the out-of-phase AC response field from a material of interest, in the strong drive field regime, can be estimated when a conventional NV Hahn echo measurement is replaced with a SIPHT Hahn echo. For example, for NV MW pulse Rabi strength $\Omega$=$2\pi \cdot$4 MHz and drive field $B_D$ =  250 $\mu$T ($\gamma B_D= 2\pi \cdot 7$ MHz), a SIPHT Hahn echo is expected to provide about 8 times the AC measurement contrast of a conventional Hahn echo.\\

\subsection{Experimental validation of SIPHT performance}
To evaluate the practical feasibility and performance of SIPHT, we now assess how well its operational conditions can be realized within our experimental setup. Specifically, we demonstrate the extent to which the drive field $B_D$ can be effectively rejected during and NV AC magnetometry measurement. We also determine how accurately the response field phase delay $\delta$ can be extracted from Hahn echo contrast measurements taken under SIPHT conditions.

\subsubsection{Effective rejection of drive field $B_{D}$}
\begin{figure}
\includegraphics[width=\linewidth]{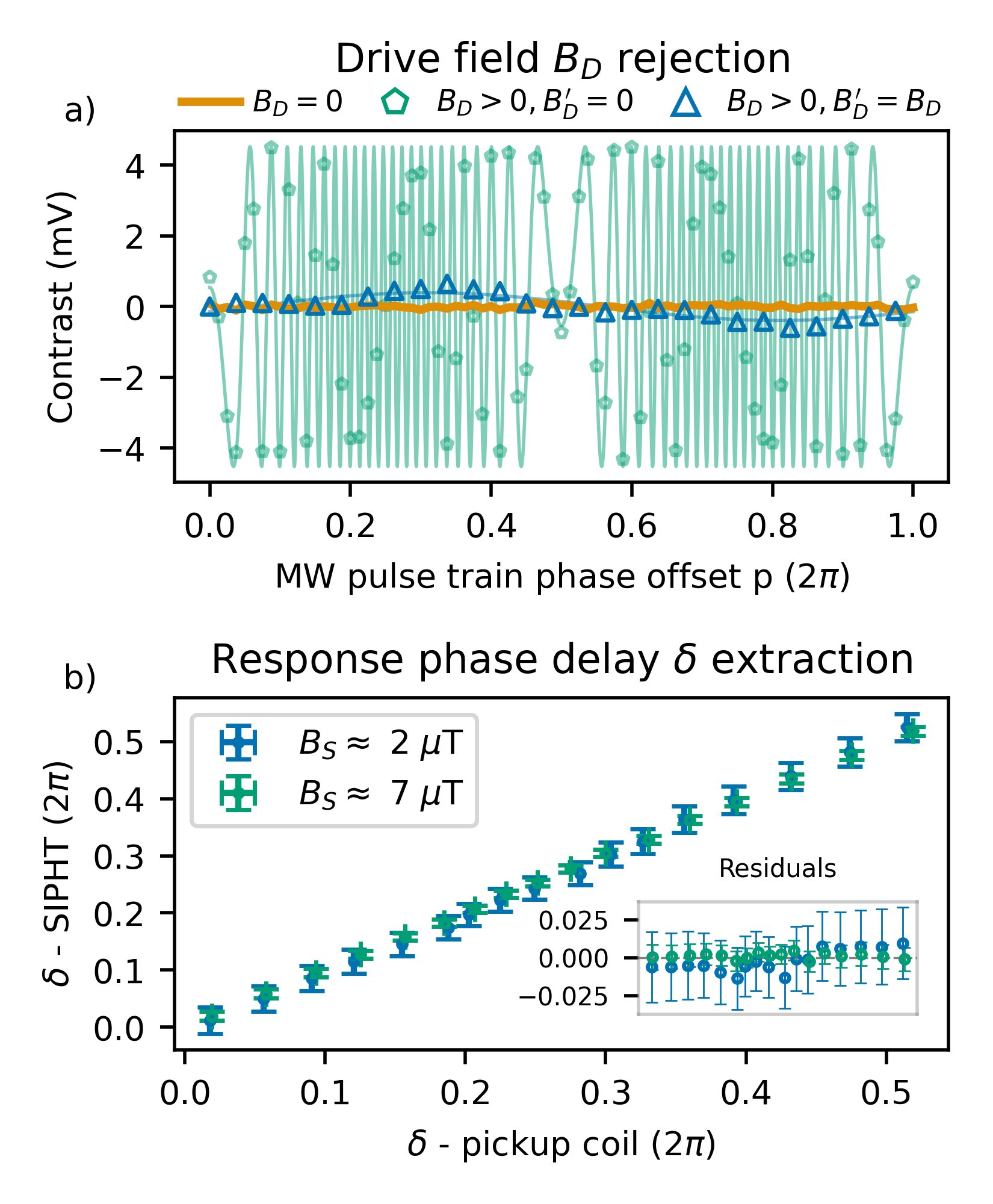}  
\caption{a) Experimental demonstration of effective drive field rejection through MW phase modulation in SIPTH DD. The phase offset $p$ of the MW pulse train is swept for 3 different conditions: a reference measurement in the absence of a drive field (orange line), a comparison measurement in the presence of a drive field without SIPHT conditions (green pentagons), and a demonstration measurement in the presence of a drive field under SIPHT conditions (blue triangles). SIPHT DD rejects $>$ 99.9\% of the drive field-induced phase for $B_D\approx102$ $\mu$T at $f_D=$ 152 kHz. b) Extraction of response field phase delay $\delta$ from an NV SIPHT Hahn echo measurement of coil-generated fields $B_S\approx$ 2 and 7 $\mu$T together with drive fields $B_D \approx$ 78 and 75 $\mu$T, respectively, at $f_D=$ 160 kHz. Good agreement is found with independent measurements performed with an inductive pickup coil over the full range of phase delays $\delta$, with residuals to a linear fit of the two measurement methods shown in the inset.}\label{fig:insensitivity}
\end{figure}

In the absence of an applied (synthetic) response signal field $B_S$, we confirm that modulation of the MW phase offset $p$ in SIPHT DD effectively cancels the influence of $B_D$ on the NV measurement contrast. This result is demonstrated by sweeping $p$ in a Hahn echo sequence, with results shown in Fig. \ref{fig:insensitivity}a, for the following three conditions:
\begin{enumerate}
    \item A reference measurement in the absence of a drive field $(B_D=0)$ and without MW pulse phase modulation $(B_D'=0)$ - orange line
    \item A comparison measurement in the presence of a drive field $(B_D>0)$, but without MW pulse phase modulation $(B_D'=0)$ - green pentagons    
    \item A demonstration measurement in the presence of a drive field $(B_D>0)$, under SIPHT conditions with MW phase modulation $(B_D'=B_D)$ - blue triangles
\end{enumerate}

SIPHT conditions can be effectively maintained for a drive field of $B_D \approx 102~\mu\text{T}$ at $f_D=$ 152 kHz, with residual phase leakage (as detected by NV AC magnetometry) suppressed to below 0.1\%. This small phase leakage is quantified by comparing the fitted phase accumulation in Eq.~\eqref{eq:phase_SIPHT} under SIPHT and conventional DD conditions; and is attributed to experimental imperfections.

\subsubsection{Extraction of response field phase delay $\delta$}
We next verify that, in the presence of both an applied drive field $B_D$ and a response field $B_S$, the response field phase delay $\delta$ can be accurately extracted from the NV AC measurement contrast under SIPHT conditions. A synthetic (coil-generated) response field with varying phase and an amplitude of either $B_S\approx$ 2 or 7 $\mu$T is added to a drive field with $f_{D}=$ 160 kHz and $B_D\approx 78$ or 75 $\mu$T, respectively. The phase delay of the response field is measured independently with both a pickup coil and an NV SIPHT Hahn echo. The response field amplitude $B_S$ and phase delay $\delta$ are extracted by fitting the observed Hahn echo contrast as a function of the phase $p$ of the MW pulse train. The fitted values of $\delta$ are plotted in Fig.~\ref{fig:insensitivity}b against the independently measured $\delta$ values obtained from the pickup coil. The inset displays the residuals, demonstrating consistent SIPHT phase extraction and uncertainties across the full range of $\delta$ values for both $B_S \approx 2~\mu$T and $7~\mu$T.

\subsection{Eddy current-induced response field}
Having established and validated the SIPHT DD protocol, we now demonstrate its application for measuring an AC response field $B_S$ generated by eddy currents in metallic structures placed in proximity to the diamond sensor. To this end, three different thin metal disks (Cu, Al, Ti) are sequentially positioned near the NV surface layer, with their plane oriented perpendicular to the (aligned) static bias field, probed NV axis, and AC drive field of $B_D\approx$ 100 $\mu$T at $f_D=$ 152 kHz. The amplitude $B_S$ and phase $\delta$ of the resulting eddy current-induced AC response field are determined from SIPTH Hahn echo measurements with $B_D'=B_D$, by sweeping the MW pulse phase offset $p$ and fitting the measured NV PL contrast to Eq. \eqref{eq:phase_SIPHT}, with calibration of accumulated NV spin phase to contrast as described in Sec. \ref{sec:methods}.\\

\begin{figure}
\includegraphics[width=\linewidth]{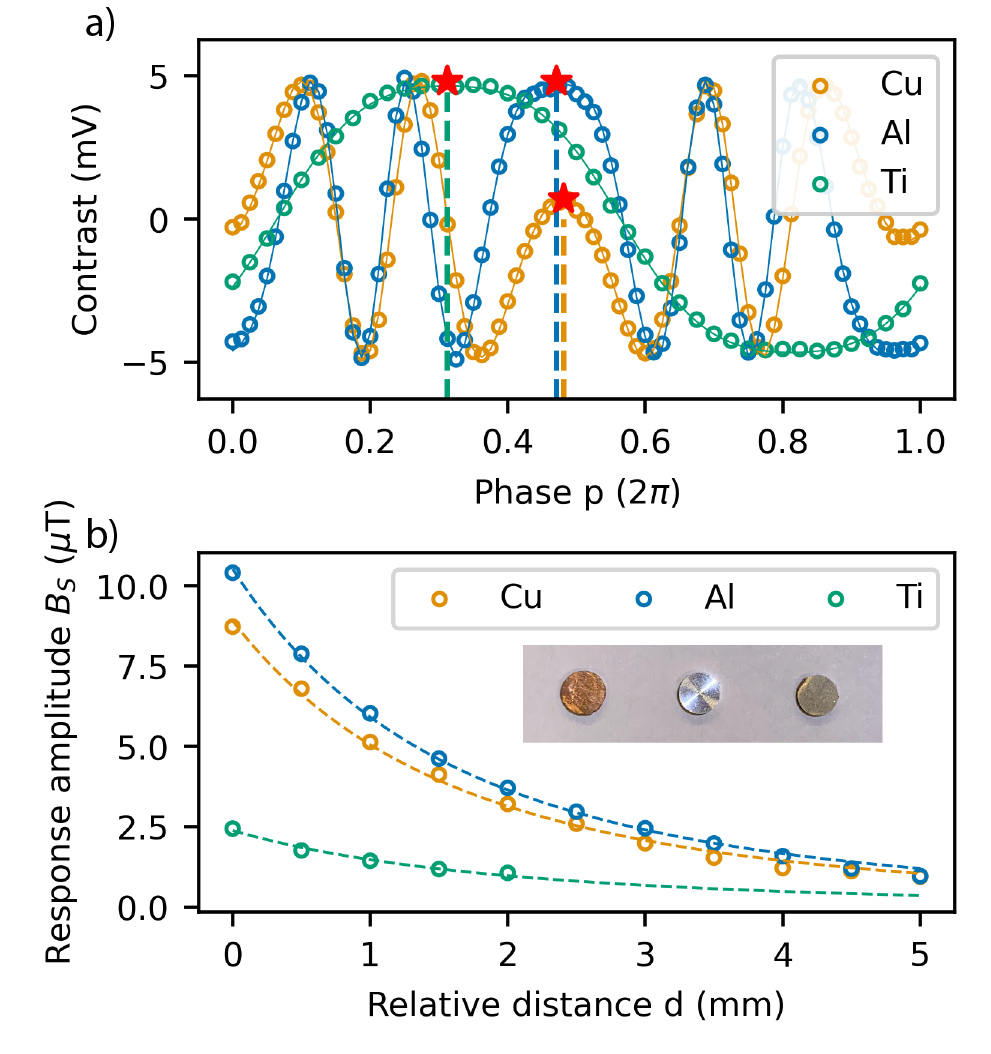}  
\caption{NV measurement of the eddy current-induced response field from Cu, Al, and Ti disks using a SIPHT Hahn echo. a) Measurement contrast as a function of the phase offset $p$ of the MW pulse train. NV spin phase accumulation is solely due to the response field induced by the eddy currents; and the phase delay $\delta$ of the response field can be directly read out from the contrast, as indicated by the red stars. b) Amplitudes $B_S$ of the eddy current-induced response fields for each metal disk (shown in inset), extracted from the fits in a), as a function of relative distance of the disk center from the NV-diamond sensor. The observed trend follows an inverse-cube scaling, consistent with a dipolar-like response field generated by eddy currents in the metal disks.}\label{fig:materials}
\end{figure}

Fig. \ref{fig:materials}a presents an example, for each metal disk, of the measured contrast and associated fits under SIPHT conditions as a function of the MW pulse phase offset $p$. As $B_D'=B_D$, NV spin phase accumulation in the AC magnetometry measurements arises solely from the response field induced by eddy currents. The response field phase delay $\delta$ can be determined from the measured contrast at values of $p$ where the contrast is symmetric, as indicated by the red stars in Fig. \ref{fig:materials}a. The extracted response field phase delays reflect the conductivity of the materials. For Cu and Al - both good conductors - the induced eddy currents generate response fields that nearly cancel the drive field, resulting in phase delays close to $p=\pi$ (i.e., $\delta=0.48\cdot 2\pi$ and $\delta=0.47\cdot 2\pi$, corresponding to 173\degree and 169\degree, respectively for the Cu and Al samples). In contrast, Ti, being a poorer conductor, produces weaker eddy currents and a smaller phase delay ($\delta=0.31\cdot 2\pi$, or 113\degree).\\

Fig. \ref{fig:materials}b shows the fitted amplitudes of $B_S$ as a function of the relative distance $d$ between the center of each metal disk and the NV-diamond sensor  (fits to the data are provided in the Supplementary Information). A functional dependence $B_S \sim 1/(d+d_0)^3$ describes the data well, with $d_0$ as an unknown offset distance. This result is consistent with the expected dipolar-like response field generated by eddy currents.

\section{Discussion and outlook\label{sec:discussion}}
As demonstrated above, SIPHT DD provides significant advantages for measuring weak AC signals in the presence of strong AC backgrounds, with wide-ranging applications. In the context of AC magnetic response measurements, the MW phase modulation used in SIPHT DD has two key features:
\begin{enumerate}
    \item MW pulse detuning is avoided in the presence of a strong AC drive field, thereby preserving NV sensitivity to the out-of-phase response field from materials of interest.
    \item NV spin phase accumulation from the AC drive field is eliminated, enabling direct readout of the response field phase delay from the measurement contrast when sweeping over the MW pulse train phase offset $p$.
\end{enumerate}
Both of these features become increasingly important when $B_D$ is large, i.e., in regimes where $B_D\gtrsim \Omega$ and $B_D\gg B_S$. As a result, SIPHT DD significantly expands the application space of NV-based AC magnetic characterization, e.g., magnetic hyperthermia of MNPs, where the drive field typically reaches several mT. SIPHT DD also extends the usable drive field range for materials previously studied with NV-based AC susceptibility, such as 2D van der Waals magnets, where previous NV measurements have been restricted to weak excitation regimes \cite{Zhang2021}.\\

In the present work, MW phase modulation is demonstrated for a sinusoidal drive field, but the approach can be adapted for other drive field waveforms. Similarly, SIPHT DD is experimentally demonstrated here using a single Hahn echo, but the technique can be easily applied to other dynamical decoupling protocols. We anticipate that the enhancement in out-of-phase field amplitude sensitivity from SIPHT DD, relative to conventional DD, will grow with increasing number of MW pulses, as NV spin phase control errors from inefficient MW pulsing in the presence of a strong AC drive field will accumulate with each additional MW pulse.\\

An additional benefit of MW phase modulation in SIPHT is enhanced control over the accumulated NV spin phase $\phi_{\text{NV}} \propto (B-B')$ via the MW phase modulation amplitude $B'$ (see Eq. \eqref{eq:phase_SIPHT}); and the resulting increased freedom to generate AC magnetometry curves for calibration purposes. By varying $B'$, one can effectively change $\phi_{NV}$, mimicking the effect of modifying the actual signal field amplitude $B$. This calibration approach is particularly useful in situations where the signal amplitude $B$ cannot be easily varied, e.g., the field from a magnetic material. SIPHT provides an alterative control parameter for phase encoding in DD sequences. Rather than measuring the accumulated phase from a fixed external field $B$, one actively nulls this phase by tuning $B'$ to match $B$, thereby determining the field amplitude in the process.

\section{Methods\label{sec:methods}}

\subsection{NV-diamond and DD sensing}
The negatively charged nitrogen-vacancy (NV) center in diamond is a point defect formed by a substitutional nitrogen atom adjacent to a lattice vacancy. Its electronic ground state is a spin triplet (S = 1), split by a zero field splitting $D \approx$ $2\pi \cdot$2.87 GHz between $m_s$ = 0 and the degenerate $m_s$ = $\pm 1$ sublevels. Applying a static bias magnetic field along the NV axis lifts the degeneracy of the $m_s$ = $\pm 1$ states via the Zeeman effect, with NV gyromagnetic ratio $\gamma \approx$ 28 MHz/mT. The spin sublevels can be manipulated using a microwave (MW) drive, by selectively addressing the allowed transitions ($m_s$ = 0 $\leftrightarrow$ $m_s$ = $-1$ in the present experiments), reducing the three-level NV ground state to an effective two-level system. Initialization and readout of this effective two-level system is done through optical excitation. 532 nm laser light repolarizes the system to the $m_s$ = 0 state and allows for spin-dependent photoluminescence (PL) readout, as the $m_s$ = 0 state emits red PL more strongly than the $m_s$ = $\pm1$ states. This combination of optical initialization, PL readout, and coherent MW control in a robust solid host makes NV centers an ideal quantum platform for room temperature magnetometry and magnetic microscopy \cite{Levine2019,Barry2020}.\\

Dynamical decoupling (DD) sequences allow measurement of narrowband (AC) magnetic fields along the NV axis by applying a train of resonant MW pulses to the NV electronic spin. After optical initialization of the NV spin, an initial $\pi/2$ MW pulse prepares the NV spin in an equal superposition of $m_s$ = $0$ and $ m_s =-1$. In the presence of a magnetic field $B(t)$, the NV spin accumulates a phase $\gamma \int B(t)dt$. A subsequent train of MW $\pi$ pulses spaced at intervals of $\tau/2$ ensures constructive NV spin phase accumulation $\phi$ in response to $B(t)$ around the DD measurement frequency of $1/\tau$, while simultaneously decoupling the NV spin from fields outside the DD measurement bandwidth, set by the inverse of the overall DD sequence time. A final MW $\pi/2$ pulse at the end of the DD sequence converts the total accumulated NV spin phase into a population difference between $m_s=0$ and $m_s=-1$, which is read out optically as a PL contrast in experiments:

\begin{equation}
    C = C_0 + C^*\cos{(\phi+\varphi)}\label{eq:cos_magn}
\end{equation}
with $\varphi$ the phase difference between the two MW $\pi/2$ pulses in the DD sequence, $C_0$ the background PL signal (not spin-state dependent), and $C^*$ the spin-state dependent PL signal. DD sequences can be applied to NV ensembles for sensitive AC magnetometry, as in the present experiments, as long as the NV properties and the optical and MW fields have sufficient spatial homogeneity over the probed region \cite{Levine2019,Barry2020}.

\subsection{Experimental details}

Experiments are performed on a (100)-cut single-crystal mm-scale diamond (Element Six Ltd., CVD-grown, 99.99 \% 12C) containing about 17 ppm substitutional 15N and an NV density of about 2.7 ppm in a layer $\approx$ 10 $\mu$m at the diamond surface. A static bias field of 5.1 mT is applied using a ring-shaped SmCo permanent magnet aligned to one of the NV orientations within the diamond. AC magnetic fields are generated by a home-built Helmholtz coil (mean diameter $\approx 55$ mm, mean height $\approx 21$ mm, 300 windings per side) with a resonance frequency near 145 kHz. The diamond and MW waveguide are mounted on the coil core with an angled surface to ensure alignment of the coil’s AC field with the NV axis used for AC magnetometry.\\

Pulse sequences are synthesized by a Keysight AWG (M8195A, dual channel with markers): channel 1 provides MW control pulses, channel 2 drives the AC coil, channel 3 provides TTL pulses to gate the acousto-optic modulator (AOM) used for control of laser illumination of the NVs, and channel 4 synchronizes the data acquisition (DAQ). For the measurements in Fig. \ref{fig:insensitivity}b, a Keysight M8190A is employed for its higher voltage amplitude resolution to efficiently generate a small AC response field on top of a large AC drive field. MW pulses are amplified (Mini-Circuits ZHL-16W-43-S+) and delivered via a circulator (Pasternack PE8401) to a $\Omega$-shaped coplanar waveguide adjacent to the NV layer of the diamond, terminated with a 50 $\Omega$ terminator. The MW drive frequency is set to the midpoint between the NV's two hyperfine transitions (split by $\approx$ 3 MHz for 15N, I=1/2).\\

NV optical excitation at 532 nm is provided by a Sprout-H laser (Lighthouse Photonics), modulated with the AOM (G\& H 3250-220). The zeroth diffraction order is blocked, and the first order (providing $\approx$ 37 mW) is focused onto the NV layer, illuminating a region of $\approx$ 40 $\mu$m. NV PL is collected using a compact aspheric lens (Thorlabs C330TMD-A), filtered with a red-pass element, and focused (f = 50 mm plano-convex) onto a photodiode (Thorlabs PDA36A2). The photodiode signal is digitized with a National Instruments DAQ (USB-6262).\\

All NV AC magnetometry measurements use a single Hahn echo ($N_{\pi}$ = 1), with AC drive field frequency between 149–160 kHz. No NV nuclear spin polarization is employed. For material eddy current measurements, circular disks of Cu (6.3 mm × 1.6 mm), Al (6.0 mm × 1.9 mm), and Ti (6.0 mm × 1.8 mm) are reproducibly mounted in proximity to the NV-diamond sensor using a non-magnetic holder with the disk normal aligned parallel to the AC drive field, the bias magnetic field, and the probed NV sensing axis.

\begin{acknowledgments}
We gratefully thank Dr. Alessandro Restelli and Nolan Ballew for their help with the construction of the AC drive coil. This work is supported by the U.S. Army Research Laboratory, under Contract No. W911NF2420143; the U.S. Army Research Office, under Grant No. W911NF2120110; the U.S. Air Force Office of Scientific Research under Grant No. FA9550-22-1-0312 as part of the MURI project Comprehensive Minimally/non-invasive Multifaceted Assessment of Nano-/microelectronic Devices (CoMMAND); and the University of Maryland Quantum Technology Center. K.E. is supported by the Belgian American Educational Foundation.
\end{acknowledgments}

\bibliography{export}% Produces the bibliography via BibTeX.

\begin{thebibliography}{23}%
\makeatletter
\providecommand \@ifxundefined [1]{%
 \@ifx{#1\undefined}
}%
\providecommand \@ifnum [1]{%
 \ifnum #1\expandafter \@firstoftwo
 \else \expandafter \@secondoftwo
 \fi
}%
\providecommand \@ifx [1]{%
 \ifx #1\expandafter \@firstoftwo
 \else \expandafter \@secondoftwo
 \fi
}%
\providecommand \natexlab [1]{#1}%
\providecommand \enquote  [1]{``#1''}%
\providecommand \bibnamefont  [1]{#1}%
\providecommand \bibfnamefont [1]{#1}%
\providecommand \citenamefont [1]{#1}%
\providecommand \href@noop [0]{\@secondoftwo}%
\providecommand \href [0]{\begingroup \@sanitize@url \@href}%
\providecommand \@href[1]{\@@startlink{#1}\@@href}%
\providecommand \@@href[1]{\endgroup#1\@@endlink}%
\providecommand \@sanitize@url [0]{\catcode `\\12\catcode `\$12\catcode
  `\&12\catcode `\#12\catcode `\^12\catcode `\_12\catcode `\%12\relax}%
\providecommand \@@startlink[1]{}%
\providecommand \@@endlink[0]{}%
\providecommand \url  [0]{\begingroup\@sanitize@url \@url }%
\providecommand \@url [1]{\endgroup\@href {#1}{\urlprefix }}%
\providecommand \urlprefix  [0]{URL }%
\providecommand \Eprint [0]{\href }%
\providecommand \doibase [0]{https://doi.org/}%
\providecommand \selectlanguage [0]{\@gobble}%
\providecommand \bibinfo  [0]{\@secondoftwo}%
\providecommand \bibfield  [0]{\@secondoftwo}%
\providecommand \translation [1]{[#1]}%
\providecommand \BibitemOpen [0]{}%
\providecommand \bibitemStop [0]{}%
\providecommand \bibitemNoStop [0]{.\EOS\space}%
\providecommand \EOS [0]{\spacefactor3000\relax}%
\providecommand \BibitemShut  [1]{\csname bibitem#1\endcsname}%
\let\auto@bib@innerbib\@empty
%</preamble>
\bibitem [{\citenamefont {Meseguer-Sánchez}\ \emph {et~al.}(2021)\citenamefont
  {Meseguer-Sánchez}, \citenamefont {Popescu}, \citenamefont {García-Muñoz},
  \citenamefont {Luetkens}, \citenamefont {Taniashvili}, \citenamefont
  {Navarro-Moratalla}, \citenamefont {Guguchia},\ and\ \citenamefont
  {Santos}}]{Meseguer-Snchez2021}%
  \BibitemOpen
  \bibfield  {author} {\bibinfo {author} {\bibfnamefont {J.}~\bibnamefont
  {Meseguer-Sánchez}}, \bibinfo {author} {\bibfnamefont {C.}~\bibnamefont
  {Popescu}}, \bibinfo {author} {\bibfnamefont {J.~L.}\ \bibnamefont
  {García-Muñoz}}, \bibinfo {author} {\bibfnamefont {H.}~\bibnamefont
  {Luetkens}}, \bibinfo {author} {\bibfnamefont {G.}~\bibnamefont
  {Taniashvili}}, \bibinfo {author} {\bibfnamefont {E.}~\bibnamefont
  {Navarro-Moratalla}}, \bibinfo {author} {\bibfnamefont {Z.}~\bibnamefont
  {Guguchia}},\ and\ \bibinfo {author} {\bibfnamefont {E.~J.}\ \bibnamefont
  {Santos}},\ }\bibfield  {title} {\bibinfo {title} {Coexistence of structural
  and magnetic phases in van der waals magnet cri3},\ }\href
  {https://www.nature.com/articles/s41467-021-26342-4} {\bibfield  {journal}
  {\bibinfo  {journal} {Nature Communications 2021}\ }\textbf {\bibinfo
  {volume} {12}},\ \bibinfo {pages} {1} (\bibinfo {year} {2021})}\BibitemShut
  {NoStop}%
\bibitem [{\citenamefont {Zhang}\ \emph {et~al.}(2021)\citenamefont {Zhang},
  \citenamefont {Wang}, \citenamefont {Tartaglia}, \citenamefont {Ding},
  \citenamefont {Gray}, \citenamefont {Burch}, \citenamefont {Tafti},\ and\
  \citenamefont {Zhou}}]{Zhang2021}%
  \BibitemOpen
  \bibfield  {author} {\bibinfo {author} {\bibfnamefont {X.-Y.}\ \bibnamefont
  {Zhang}}, \bibinfo {author} {\bibfnamefont {Y.-X.}\ \bibnamefont {Wang}},
  \bibinfo {author} {\bibfnamefont {T.~A.}\ \bibnamefont {Tartaglia}}, \bibinfo
  {author} {\bibfnamefont {T.}~\bibnamefont {Ding}}, \bibinfo {author}
  {\bibfnamefont {M.~J.}\ \bibnamefont {Gray}}, \bibinfo {author}
  {\bibfnamefont {K.~S.}\ \bibnamefont {Burch}}, \bibinfo {author}
  {\bibfnamefont {F.}~\bibnamefont {Tafti}},\ and\ \bibinfo {author}
  {\bibfnamefont {B.~B.}\ \bibnamefont {Zhou}},\ }\bibfield  {title} {\bibinfo
  {title} {ac susceptometry of 2d van der waals magnets enabled by the coherent
  control of quantum sensors},\ }\href@noop {} {\bibfield  {journal} {\bibinfo
  {journal} {PRX QUANTUM}\ }\textbf {\bibinfo {volume} {2}},\ \bibinfo {pages}
  {030352} (\bibinfo {year} {2021})}\BibitemShut {NoStop}%
\bibitem [{\citenamefont {Tschudin}\ \emph {et~al.}(2024)\citenamefont
  {Tschudin}, \citenamefont {Broadway}, \citenamefont {Siegwolf}, \citenamefont
  {Schrader}, \citenamefont {Telford}, \citenamefont {Gross}, \citenamefont
  {Cox}, \citenamefont {Dubois}, \citenamefont {Chica}, \citenamefont
  {Rama-Eiroa}, \citenamefont {Santos}, \citenamefont {Poggio}, \citenamefont
  {Ziebel}, \citenamefont {Dean}, \citenamefont {Roy},\ and\ \citenamefont
  {Maletinsky}}]{Tschudin2024}%
  \BibitemOpen
  \bibfield  {author} {\bibinfo {author} {\bibfnamefont {M.~A.}\ \bibnamefont
  {Tschudin}}, \bibinfo {author} {\bibfnamefont {D.~A.}\ \bibnamefont
  {Broadway}}, \bibinfo {author} {\bibfnamefont {P.}~\bibnamefont {Siegwolf}},
  \bibinfo {author} {\bibfnamefont {C.}~\bibnamefont {Schrader}}, \bibinfo
  {author} {\bibfnamefont {E.~J.}\ \bibnamefont {Telford}}, \bibinfo {author}
  {\bibfnamefont {B.}~\bibnamefont {Gross}}, \bibinfo {author} {\bibfnamefont
  {J.}~\bibnamefont {Cox}}, \bibinfo {author} {\bibfnamefont {A.~E.}\
  \bibnamefont {Dubois}}, \bibinfo {author} {\bibfnamefont {D.~G.}\
  \bibnamefont {Chica}}, \bibinfo {author} {\bibfnamefont {R.}~\bibnamefont
  {Rama-Eiroa}}, \bibinfo {author} {\bibfnamefont {E.~J.~G.}\ \bibnamefont
  {Santos}}, \bibinfo {author} {\bibfnamefont {M.}~\bibnamefont {Poggio}},
  \bibinfo {author} {\bibfnamefont {M.~E.}\ \bibnamefont {Ziebel}}, \bibinfo
  {author} {\bibfnamefont {C.~R.}\ \bibnamefont {Dean}}, \bibinfo {author}
  {\bibfnamefont {X.}~\bibnamefont {Roy}},\ and\ \bibinfo {author}
  {\bibfnamefont {P.}~\bibnamefont {Maletinsky}},\ }\bibfield  {title}
  {\bibinfo {title} {Imaging nanomagnetism and magnetic phase transitions in
  atomically thin crsbr},\ }\href
  {https://www.nature.com/articles/s41467-024-49717-9} {\bibfield  {journal}
  {\bibinfo  {journal} {Nature Communications 2024}\ }\textbf {\bibinfo
  {volume} {15}},\ \bibinfo {pages} {1} (\bibinfo {year} {2024})}\BibitemShut
  {NoStop}%
\bibitem [{\citenamefont {García-Martín}\ \emph {et~al.}(2011)\citenamefont
  {García-Martín}, \citenamefont {Gómez-Gil},\ and\ \citenamefont
  {Vázquez-Sánchez}}]{Garca-Martn2011}%
  \BibitemOpen
  \bibfield  {author} {\bibinfo {author} {\bibfnamefont {J.}~\bibnamefont
  {García-Martín}}, \bibinfo {author} {\bibfnamefont {J.}~\bibnamefont
  {Gómez-Gil}},\ and\ \bibinfo {author} {\bibfnamefont {E.}~\bibnamefont
  {Vázquez-Sánchez}},\ }\bibfield  {title} {\bibinfo {title} {Non-destructive
  techniques based on eddy current testing},\ }\href
  {https://www.mdpi.com/1424-8220/11/3/2525/htm
  https://www.mdpi.com/1424-8220/11/3/2525} {\bibfield  {journal} {\bibinfo
  {journal} {Sensors 2011, Vol. 11, Pages 2525-2565}\ }\textbf {\bibinfo
  {volume} {11}},\ \bibinfo {pages} {2525} (\bibinfo {year}
  {2011})}\BibitemShut {NoStop}%
\bibitem [{\citenamefont {Wickenbrock}\ \emph {et~al.}(2016)\citenamefont
  {Wickenbrock}, \citenamefont {Leefer}, \citenamefont {Blanchard},\ and\
  \citenamefont {Budker}}]{Wickenbrock2016}%
  \BibitemOpen
  \bibfield  {author} {\bibinfo {author} {\bibfnamefont {A.}~\bibnamefont
  {Wickenbrock}}, \bibinfo {author} {\bibfnamefont {N.}~\bibnamefont {Leefer}},
  \bibinfo {author} {\bibfnamefont {J.~W.}\ \bibnamefont {Blanchard}},\ and\
  \bibinfo {author} {\bibfnamefont {D.}~\bibnamefont {Budker}},\ }\bibfield
  {title} {\bibinfo {title} {Eddy current imaging with an atomic
  radio-frequency magnetometer},\ }\href
  {/aip/apl/article/108/18/183507/593896/Eddy-current-imaging-with-an-atomic-radio}
  {\bibfield  {journal} {\bibinfo  {journal} {Applied Physics Letters}\
  }\textbf {\bibinfo {volume} {108}},\ \bibinfo {pages} {46} (\bibinfo {year}
  {2016})}\BibitemShut {NoStop}%
\bibitem [{\citenamefont {Chatzidrosos}\ \emph {et~al.}(2019)\citenamefont
  {Chatzidrosos}, \citenamefont {Wickenbrock}, \citenamefont {Bougas},
  \citenamefont {Zheng}, \citenamefont {Tretiak}, \citenamefont {Yang},\ and\
  \citenamefont {Budker}}]{Chatzidrosos2019}%
  \BibitemOpen
  \bibfield  {author} {\bibinfo {author} {\bibfnamefont {G.}~\bibnamefont
  {Chatzidrosos}}, \bibinfo {author} {\bibfnamefont {A.}~\bibnamefont
  {Wickenbrock}}, \bibinfo {author} {\bibfnamefont {L.}~\bibnamefont {Bougas}},
  \bibinfo {author} {\bibfnamefont {H.}~\bibnamefont {Zheng}}, \bibinfo
  {author} {\bibfnamefont {O.}~\bibnamefont {Tretiak}}, \bibinfo {author}
  {\bibfnamefont {Y.}~\bibnamefont {Yang}},\ and\ \bibinfo {author}
  {\bibfnamefont {D.}~\bibnamefont {Budker}},\ }\bibfield  {title} {\bibinfo
  {title} {Eddy-current imaging with nitrogen-vacancy centers in diamond},\
  }\href@noop {} {\bibfield  {journal} {\bibinfo  {journal} {Phys. Rev. Appl.}\
  }\textbf {\bibinfo {volume} {11}},\ \bibinfo {pages} {14060} (\bibinfo {year}
  {2019})}\BibitemShut {NoStop}%
\bibitem [{\citenamefont {Rosensweig}(2002)}]{Rosensweig2002}%
  \BibitemOpen
  \bibfield  {author} {\bibinfo {author} {\bibfnamefont {R.~E.}\ \bibnamefont
  {Rosensweig}},\ }\bibfield  {title} {\bibinfo {title} {Heating magnetic fluid
  with alternating magnetic field},\ }\href@noop {} {\bibfield  {journal}
  {\bibinfo  {journal} {Journal of Magnetism and Magnetic Materials}\ }\textbf
  {\bibinfo {volume} {252}},\ \bibinfo {pages} {370} (\bibinfo {year}
  {2002})}\BibitemShut {NoStop}%
\bibitem [{\citenamefont {Rodrigues}\ \emph {et~al.}(2020)\citenamefont
  {Rodrigues}, \citenamefont {Capistrano},\ and\ \citenamefont
  {Bakuzis}}]{Rodrigues2020}%
  \BibitemOpen
  \bibfield  {author} {\bibinfo {author} {\bibfnamefont {H.~F.}\ \bibnamefont
  {Rodrigues}}, \bibinfo {author} {\bibfnamefont {G.}~\bibnamefont
  {Capistrano}},\ and\ \bibinfo {author} {\bibfnamefont {A.~F.}\ \bibnamefont
  {Bakuzis}},\ }\bibfield  {title} {\bibinfo {title} {In vivo magnetic
  nanoparticle hyperthermia: a review on preclinical studies, low-field
  nano-heaters, noninvasive thermometry and computer simulations for treatment
  planning},\ }\href@noop {} {\bibfield  {journal} {\bibinfo  {journal}
  {International Journal of Hyperthermia}\ }\textbf {\bibinfo {volume} {37}},\
  \bibinfo {pages} {76} (\bibinfo {year} {2020})}\BibitemShut {NoStop}%
\bibitem [{\citenamefont {Rubia-Rodríguez}\ \emph {et~al.}(2021)\citenamefont
  {Rubia-Rodríguez}, \citenamefont {Santana-Otero}, \citenamefont {Spassov},
  \citenamefont {Tombácz}, \citenamefont {Johansson}, \citenamefont {De},
  \citenamefont {Presa}, \citenamefont {Teran}, \citenamefont {Morales},
  \citenamefont {Veintemillas-Verdaguer}, \citenamefont {Thanh}, \citenamefont
  {Besenhard}, \citenamefont {Wilhelm}, \citenamefont {Gazeau}, \citenamefont
  {Harmer}, \citenamefont {Mayes}, \citenamefont {Manshian}, \citenamefont
  {Soenen}, \citenamefont {Gu}, \citenamefont {Ángel Millán}, \citenamefont
  {Efthimiadou}, \citenamefont {Gaudet}, \citenamefont {Goodwill},
  \citenamefont {Mansfield}, \citenamefont {Steinhoff}, \citenamefont {Wells},
  \citenamefont {Wiekhorst},\ and\ \citenamefont
  {Ortega}}]{Rubia-Rodrguez2021}%
  \BibitemOpen
  \bibfield  {author} {\bibinfo {author} {\bibfnamefont {I.}~\bibnamefont
  {Rubia-Rodríguez}}, \bibinfo {author} {\bibfnamefont {A.}~\bibnamefont
  {Santana-Otero}}, \bibinfo {author} {\bibfnamefont {S.}~\bibnamefont
  {Spassov}}, \bibinfo {author} {\bibfnamefont {E.}~\bibnamefont {Tombácz}},
  \bibinfo {author} {\bibfnamefont {C.}~\bibnamefont {Johansson}}, \bibinfo
  {author} {\bibfnamefont {P.}~\bibnamefont {De}}, \bibinfo {author}
  {\bibfnamefont {L.}~\bibnamefont {Presa}}, \bibinfo {author} {\bibfnamefont
  {F.~J.}\ \bibnamefont {Teran}}, \bibinfo {author} {\bibfnamefont {M.~D.~P.}\
  \bibnamefont {Morales}}, \bibinfo {author} {\bibfnamefont {S.}~\bibnamefont
  {Veintemillas-Verdaguer}}, \bibinfo {author} {\bibfnamefont {N.~T.~K.}\
  \bibnamefont {Thanh}}, \bibinfo {author} {\bibfnamefont {M.~O.}\ \bibnamefont
  {Besenhard}}, \bibinfo {author} {\bibfnamefont {C.}~\bibnamefont {Wilhelm}},
  \bibinfo {author} {\bibfnamefont {F.}~\bibnamefont {Gazeau}}, \bibinfo
  {author} {\bibfnamefont {Q.}~\bibnamefont {Harmer}}, \bibinfo {author}
  {\bibfnamefont {E.}~\bibnamefont {Mayes}}, \bibinfo {author} {\bibfnamefont
  {B.~B.}\ \bibnamefont {Manshian}}, \bibinfo {author} {\bibfnamefont {S.~J.}\
  \bibnamefont {Soenen}}, \bibinfo {author} {\bibfnamefont {Y.}~\bibnamefont
  {Gu}}, \bibinfo {author} {\bibnamefont {Ángel Millán}}, \bibinfo {author}
  {\bibfnamefont {E.~K.}\ \bibnamefont {Efthimiadou}}, \bibinfo {author}
  {\bibfnamefont {J.}~\bibnamefont {Gaudet}}, \bibinfo {author} {\bibfnamefont
  {P.}~\bibnamefont {Goodwill}}, \bibinfo {author} {\bibfnamefont
  {J.}~\bibnamefont {Mansfield}}, \bibinfo {author} {\bibfnamefont
  {U.}~\bibnamefont {Steinhoff}}, \bibinfo {author} {\bibfnamefont
  {J.}~\bibnamefont {Wells}}, \bibinfo {author} {\bibfnamefont
  {F.}~\bibnamefont {Wiekhorst}},\ and\ \bibinfo {author} {\bibfnamefont
  {D.}~\bibnamefont {Ortega}},\ }\bibfield  {title} {\bibinfo {title} {Whither
  magnetic hyperthermia? a tentative roadmap},\ }\href
  {https://doi.org/10.3390/ma14040706} {\  (\bibinfo {year}
  {2021})}\BibitemShut {NoStop}%
\bibitem [{\citenamefont {Cabrera}\ \emph {et~al.}(2018)\citenamefont
  {Cabrera}, \citenamefont {Coene}, \citenamefont {Leliaert}, \citenamefont
  {Arte}, \citenamefont {Dupre}, \citenamefont {Telling},\ and\ \citenamefont
  {Teran}}]{Cabrera2018}%
  \BibitemOpen
  \bibfield  {author} {\bibinfo {author} {\bibfnamefont {D.}~\bibnamefont
  {Cabrera}}, \bibinfo {author} {\bibfnamefont {A.}~\bibnamefont {Coene}},
  \bibinfo {author} {\bibfnamefont {J.}~\bibnamefont {Leliaert}}, \bibinfo
  {author} {\bibfnamefont {E.~J.}\ \bibnamefont {Arte}}, \bibinfo {author}
  {\bibfnamefont {L.}~\bibnamefont {Dupre}}, \bibinfo {author} {\bibfnamefont
  {N.~D.}\ \bibnamefont {Telling}},\ and\ \bibinfo {author} {\bibfnamefont
  {F.~J.}\ \bibnamefont {Teran}},\ }\bibfield  {title} {\bibinfo {title}
  {Dynamical magnetic response of iron oxide nanoparticles inside live cells},\
  }\href {www.acsnano.org} {\bibfield  {journal} {\bibinfo  {journal} {ACS
  Nano}\ }\textbf {\bibinfo {volume} {12}},\ \bibinfo {pages} {2741} (\bibinfo
  {year} {2018})}\BibitemShut {NoStop}%
\bibitem [{\citenamefont {Gavilán}\ \emph {et~al.}(2021)\citenamefont
  {Gavilán}, \citenamefont {Avugadda}, \citenamefont {Fernández-Cabada},
  \citenamefont {Soni}, \citenamefont {Cassani}, \citenamefont {Mai},
  \citenamefont {Chantrell},\ and\ \citenamefont {Pellegrino}}]{Gaviln2021}%
  \BibitemOpen
  \bibfield  {author} {\bibinfo {author} {\bibfnamefont {H.}~\bibnamefont
  {Gavilán}}, \bibinfo {author} {\bibfnamefont {S.~K.}\ \bibnamefont
  {Avugadda}}, \bibinfo {author} {\bibfnamefont {T.}~\bibnamefont
  {Fernández-Cabada}}, \bibinfo {author} {\bibfnamefont {N.}~\bibnamefont
  {Soni}}, \bibinfo {author} {\bibfnamefont {M.}~\bibnamefont {Cassani}},
  \bibinfo {author} {\bibfnamefont {B.~T.}\ \bibnamefont {Mai}}, \bibinfo
  {author} {\bibfnamefont {R.}~\bibnamefont {Chantrell}},\ and\ \bibinfo
  {author} {\bibfnamefont {T.}~\bibnamefont {Pellegrino}},\ }\bibfield  {title}
  {\bibinfo {title} {Magnetic nanoparticles and clusters for magnetic
  hyperthermia: optimizing their heat performance and developing combinatorial
  therapies to tackle cancer},\ }\href
  {https://pubs.rsc.org/en/content/articlehtml/2021/cs/d1cs00427a
  https://pubs.rsc.org/en/content/articlelanding/2021/cs/d1cs00427a} {\bibfield
   {journal} {\bibinfo  {journal} {Chemical Society Reviews}\ }\textbf
  {\bibinfo {volume} {50}},\ \bibinfo {pages} {11614} (\bibinfo {year}
  {2021})}\BibitemShut {NoStop}%
\bibitem [{\citenamefont {Sharifabad}\ \emph {et~al.}(2024)\citenamefont
  {Sharifabad}, \citenamefont {Soucaille}, \citenamefont {Wang}, \citenamefont
  {Rotherham}, \citenamefont {Loughran}, \citenamefont {Everett}, \citenamefont
  {Cabrera}, \citenamefont {Yang}, \citenamefont {Hicken},\ and\ \citenamefont
  {Telling}}]{Sharifabad2024}%
  \BibitemOpen
  \bibfield  {author} {\bibinfo {author} {\bibfnamefont {M.~E.}\ \bibnamefont
  {Sharifabad}}, \bibinfo {author} {\bibfnamefont {R.}~\bibnamefont
  {Soucaille}}, \bibinfo {author} {\bibfnamefont {X.}~\bibnamefont {Wang}},
  \bibinfo {author} {\bibfnamefont {M.}~\bibnamefont {Rotherham}}, \bibinfo
  {author} {\bibfnamefont {T.}~\bibnamefont {Loughran}}, \bibinfo {author}
  {\bibfnamefont {J.}~\bibnamefont {Everett}}, \bibinfo {author} {\bibfnamefont
  {D.}~\bibnamefont {Cabrera}}, \bibinfo {author} {\bibfnamefont
  {Y.}~\bibnamefont {Yang}}, \bibinfo {author} {\bibfnamefont {R.}~\bibnamefont
  {Hicken}},\ and\ \bibinfo {author} {\bibfnamefont {N.}~\bibnamefont
  {Telling}},\ }\bibfield  {title} {\bibinfo {title} {Optical microscopy using
  the faraday effect reveals in situ magnetization dynamics of magnetic
  nanoparticles in biological samples},\ }\href
  {https://pubs.acs.org/doi/full/10.1021/acsnano.3c08955} {\bibfield  {journal}
  {\bibinfo  {journal} {ACS Nano}\ } (\bibinfo {year} {2024})}\BibitemShut
  {NoStop}%
\bibitem [{\citenamefont {Carrey}\ \emph {et~al.}(2011)\citenamefont {Carrey},
  \citenamefont {Mehdaoui},\ and\ \citenamefont {Respaud}}]{Carrey2011}%
  \BibitemOpen
  \bibfield  {author} {\bibinfo {author} {\bibfnamefont {J.}~\bibnamefont
  {Carrey}}, \bibinfo {author} {\bibfnamefont {B.}~\bibnamefont {Mehdaoui}},\
  and\ \bibinfo {author} {\bibfnamefont {M.}~\bibnamefont {Respaud}},\
  }\bibfield  {title} {\bibinfo {title} {Simple models for dynamic hysteresis
  loop calculations of magnetic single-domain nanoparticles: Application to
  magnetic hyperthermia optimization},\ }\href
  {https://doi.org/10.1063/1.3551582} {\bibfield  {journal} {\bibinfo
  {journal} {J. Appl. Phys}\ }\textbf {\bibinfo {volume} {109}},\ \bibinfo
  {pages} {83921} (\bibinfo {year} {2011})}\BibitemShut {NoStop}%
\bibitem [{\citenamefont {Dennis}\ and\ \citenamefont
  {Ivkov}(2013)}]{Dennis2013}%
  \BibitemOpen
  \bibfield  {author} {\bibinfo {author} {\bibfnamefont {C.~L.}\ \bibnamefont
  {Dennis}}\ and\ \bibinfo {author} {\bibfnamefont {R.}~\bibnamefont {Ivkov}},\
  }\href@noop {} {\bibinfo {title} {Physics of heat generation using magnetic
  nanoparticles for hyperthermia}} (\bibinfo {year} {2013})\BibitemShut
  {NoStop}%
\bibitem [{\citenamefont {Everaert}\ \emph {et~al.}(2024)\citenamefont
  {Everaert}, \citenamefont {Eberbeck}, \citenamefont {Körber}, \citenamefont
  {Radon}, \citenamefont {Waeyenberge}, \citenamefont {Leliaert},\ and\
  \citenamefont {Wiekhorst}}]{Everaert2024}%
  \BibitemOpen
  \bibfield  {author} {\bibinfo {author} {\bibfnamefont {K.}~\bibnamefont
  {Everaert}}, \bibinfo {author} {\bibfnamefont {D.}~\bibnamefont {Eberbeck}},
  \bibinfo {author} {\bibfnamefont {R.}~\bibnamefont {Körber}}, \bibinfo
  {author} {\bibfnamefont {P.}~\bibnamefont {Radon}}, \bibinfo {author}
  {\bibfnamefont {B.~V.}\ \bibnamefont {Waeyenberge}}, \bibinfo {author}
  {\bibfnamefont {J.}~\bibnamefont {Leliaert}},\ and\ \bibinfo {author}
  {\bibfnamefont {F.}~\bibnamefont {Wiekhorst}},\ }\bibfield  {title} {\bibinfo
  {title} {Comparing magnetization fluctuations and dissipation in suspended
  magnetic nanoparticle ensembles},\ }\href@noop {} {\bibfield  {journal}
  {\bibinfo  {journal} {IEEE Transactions on Magnetics}\ }\textbf {\bibinfo
  {volume} {60}} (\bibinfo {year} {2024})}\BibitemShut {NoStop}%
\bibitem [{\citenamefont {Taylor}\ \emph {et~al.}(2008)\citenamefont {Taylor},
  \citenamefont {Cappellaro}, \citenamefont {Childress}, \citenamefont {Jiang},
  \citenamefont {Budker}, \citenamefont {Hemmer}, \citenamefont {Yacoby},
  \citenamefont {Walsworth},\ and\ \citenamefont {Lukin}}]{Taylor2008}%
  \BibitemOpen
  \bibfield  {author} {\bibinfo {author} {\bibfnamefont {J.~M.}\ \bibnamefont
  {Taylor}}, \bibinfo {author} {\bibfnamefont {P.}~\bibnamefont {Cappellaro}},
  \bibinfo {author} {\bibfnamefont {L.}~\bibnamefont {Childress}}, \bibinfo
  {author} {\bibfnamefont {L.}~\bibnamefont {Jiang}}, \bibinfo {author}
  {\bibfnamefont {D.}~\bibnamefont {Budker}}, \bibinfo {author} {\bibfnamefont
  {P.~R.}\ \bibnamefont {Hemmer}}, \bibinfo {author} {\bibfnamefont
  {A.}~\bibnamefont {Yacoby}}, \bibinfo {author} {\bibfnamefont
  {R.}~\bibnamefont {Walsworth}},\ and\ \bibinfo {author} {\bibfnamefont
  {M.~D.}\ \bibnamefont {Lukin}},\ }\bibfield  {title} {\bibinfo {title}
  {High-sensitivity diamond magnetometer with nanoscale resolution},\ }\href
  {https://www.nature.com/articles/nphys1075} {\bibfield  {journal} {\bibinfo
  {journal} {Nature Physics}\ }\textbf {\bibinfo {volume} {4}},\ \bibinfo
  {pages} {810} (\bibinfo {year} {2008})}\BibitemShut {NoStop}%
\bibitem [{\citenamefont {Rovny}\ \emph {et~al.}(2024)\citenamefont {Rovny},
  \citenamefont {Gopalakrishnan}, \citenamefont {Jayich}, \citenamefont
  {Maletinsky}, \citenamefont {Demler},\ and\ \citenamefont
  {Leon}}]{Rovny2024}%
  \BibitemOpen
  \bibfield  {author} {\bibinfo {author} {\bibfnamefont {J.}~\bibnamefont
  {Rovny}}, \bibinfo {author} {\bibfnamefont {S.}~\bibnamefont
  {Gopalakrishnan}}, \bibinfo {author} {\bibfnamefont {A.~C.~B.}\ \bibnamefont
  {Jayich}}, \bibinfo {author} {\bibfnamefont {P.}~\bibnamefont {Maletinsky}},
  \bibinfo {author} {\bibfnamefont {E.}~\bibnamefont {Demler}},\ and\ \bibinfo
  {author} {\bibfnamefont {N.~P.~D.}\ \bibnamefont {Leon}},\ }\bibfield
  {title} {\bibinfo {title} {Nanoscale diamond quantum sensors for many-body
  physics},\ }\href@noop {} {\bibfield  {journal} {\bibinfo  {journal} {Nature
  Reviews Physics}\ }\textbf {\bibinfo {volume} {6}},\ \bibinfo {pages} {753}
  (\bibinfo {year} {2024})}\BibitemShut {NoStop}%
\bibitem [{\citenamefont {Sage}\ \emph {et~al.}(2013)\citenamefont {Sage},
  \citenamefont {Arai}, \citenamefont {Glenn}, \citenamefont {DeVience},
  \citenamefont {Pham}, \citenamefont {Rahn-Lee}, \citenamefont {Lukin},
  \citenamefont {Yacoby}, \citenamefont {Komeili},\ and\ \citenamefont
  {Walsworth}}]{LeSage2013}%
  \BibitemOpen
  \bibfield  {author} {\bibinfo {author} {\bibfnamefont {D.~L.}\ \bibnamefont
  {Sage}}, \bibinfo {author} {\bibfnamefont {K.}~\bibnamefont {Arai}}, \bibinfo
  {author} {\bibfnamefont {D.~R.}\ \bibnamefont {Glenn}}, \bibinfo {author}
  {\bibfnamefont {S.~J.}\ \bibnamefont {DeVience}}, \bibinfo {author}
  {\bibfnamefont {L.~M.}\ \bibnamefont {Pham}}, \bibinfo {author}
  {\bibfnamefont {L.}~\bibnamefont {Rahn-Lee}}, \bibinfo {author}
  {\bibfnamefont {M.~D.}\ \bibnamefont {Lukin}}, \bibinfo {author}
  {\bibfnamefont {A.}~\bibnamefont {Yacoby}}, \bibinfo {author} {\bibfnamefont
  {A.}~\bibnamefont {Komeili}},\ and\ \bibinfo {author} {\bibfnamefont {R.~L.}\
  \bibnamefont {Walsworth}},\ }\bibfield  {title} {\bibinfo {title} {Optical
  magnetic imaging of living cells},\ }\href@noop {} {\bibfield  {journal}
  {\bibinfo  {journal} {Nature}\ }\textbf {\bibinfo {volume} {496}},\ \bibinfo
  {pages} {486} (\bibinfo {year} {2013})}\BibitemShut {NoStop}%
\bibitem [{\citenamefont {Glenn}\ \emph {et~al.}(2015)\citenamefont {Glenn},
  \citenamefont {Lee}, \citenamefont {Park}, \citenamefont {Weissleder},
  \citenamefont {Yacoby}, \citenamefont {Lukin}, \citenamefont {Lee},
  \citenamefont {Walsworth},\ and\ \citenamefont {Connolly}}]{Glenn2015}%
  \BibitemOpen
  \bibfield  {author} {\bibinfo {author} {\bibfnamefont {D.~R.}\ \bibnamefont
  {Glenn}}, \bibinfo {author} {\bibfnamefont {K.}~\bibnamefont {Lee}}, \bibinfo
  {author} {\bibfnamefont {H.}~\bibnamefont {Park}}, \bibinfo {author}
  {\bibfnamefont {R.}~\bibnamefont {Weissleder}}, \bibinfo {author}
  {\bibfnamefont {A.}~\bibnamefont {Yacoby}}, \bibinfo {author} {\bibfnamefont
  {M.~D.}\ \bibnamefont {Lukin}}, \bibinfo {author} {\bibfnamefont
  {H.}~\bibnamefont {Lee}}, \bibinfo {author} {\bibfnamefont {R.~L.}\
  \bibnamefont {Walsworth}},\ and\ \bibinfo {author} {\bibfnamefont {C.~B.}\
  \bibnamefont {Connolly}},\ }\bibfield  {title} {\bibinfo {title} {Single-cell
  magnetic imaging using a quantum diamond microscope},\ }\href
  {https://www.nature.com/articles/nmeth.3449} {\bibfield  {journal} {\bibinfo
  {journal} {Nature Methods}\ }\textbf {\bibinfo {volume} {12}},\ \bibinfo
  {pages} {736} (\bibinfo {year} {2015})}\BibitemShut {NoStop}%
\bibitem [{\citenamefont {Nolte}\ \emph {et~al.}(2014)\citenamefont {Nolte},
  \citenamefont {Schlipf}, \citenamefont {Ternes}, \citenamefont {Reinhard},
  \citenamefont {Kern},\ and\ \citenamefont {Wrachtrup}}]{SchaferNolte2014}%
  \BibitemOpen
  \bibfield  {author} {\bibinfo {author} {\bibfnamefont {E.~S.}\ \bibnamefont
  {Nolte}}, \bibinfo {author} {\bibfnamefont {L.}~\bibnamefont {Schlipf}},
  \bibinfo {author} {\bibfnamefont {M.}~\bibnamefont {Ternes}}, \bibinfo
  {author} {\bibfnamefont {F.}~\bibnamefont {Reinhard}}, \bibinfo {author}
  {\bibfnamefont {K.}~\bibnamefont {Kern}},\ and\ \bibinfo {author}
  {\bibfnamefont {J.}~\bibnamefont {Wrachtrup}},\ }\bibfield  {title} {\bibinfo
  {title} {Tracking temperature-dependent relaxation times of ferritin
  nanomagnets with a wideband quantum spectrometer},\ }\href@noop {} {\bibfield
   {journal} {\bibinfo  {journal} {Physical Review Letters}\ }\textbf {\bibinfo
  {volume} {113}},\ \bibinfo {pages} {217204} (\bibinfo {year}
  {2014})}\BibitemShut {NoStop}%
\bibitem [{\citenamefont {Richards}\ \emph {et~al.}(2025)\citenamefont
  {Richards}, \citenamefont {Ristoff}, \citenamefont {Smits}, \citenamefont
  {Perez}, \citenamefont {Fescenko}, \citenamefont {Aiello}, \citenamefont
  {Hubert}, \citenamefont {Silani}, \citenamefont {Mosavian}, \citenamefont
  {Ziabari}, \citenamefont {Berzins}, \citenamefont {Damron}, \citenamefont
  {Kehayias}, \citenamefont {Egbebunmi}, \citenamefont {Shield}, \citenamefont
  {Huber}, \citenamefont {Mounce}, \citenamefont {Lilly}, \citenamefont
  {Karaulanov}, \citenamefont {Jarmola}, \citenamefont {Laraoui},\ and\
  \citenamefont {Acosta}}]{Richards2025}%
  \BibitemOpen
  \bibfield  {author} {\bibinfo {author} {\bibfnamefont {B.~A.}\ \bibnamefont
  {Richards}}, \bibinfo {author} {\bibfnamefont {N.}~\bibnamefont {Ristoff}},
  \bibinfo {author} {\bibfnamefont {J.}~\bibnamefont {Smits}}, \bibinfo
  {author} {\bibfnamefont {A.~J.}\ \bibnamefont {Perez}}, \bibinfo {author}
  {\bibfnamefont {I.}~\bibnamefont {Fescenko}}, \bibinfo {author}
  {\bibfnamefont {M.~D.}\ \bibnamefont {Aiello}}, \bibinfo {author}
  {\bibfnamefont {F.}~\bibnamefont {Hubert}}, \bibinfo {author} {\bibfnamefont
  {Y.}~\bibnamefont {Silani}}, \bibinfo {author} {\bibfnamefont
  {N.}~\bibnamefont {Mosavian}}, \bibinfo {author} {\bibfnamefont {M.~S.}\
  \bibnamefont {Ziabari}}, \bibinfo {author} {\bibfnamefont {A.}~\bibnamefont
  {Berzins}}, \bibinfo {author} {\bibfnamefont {J.~T.}\ \bibnamefont {Damron}},
  \bibinfo {author} {\bibfnamefont {P.}~\bibnamefont {Kehayias}}, \bibinfo
  {author} {\bibfnamefont {D.}~\bibnamefont {Egbebunmi}}, \bibinfo {author}
  {\bibfnamefont {J.~E.}\ \bibnamefont {Shield}}, \bibinfo {author}
  {\bibfnamefont {D.~L.}\ \bibnamefont {Huber}}, \bibinfo {author}
  {\bibfnamefont {A.~M.}\ \bibnamefont {Mounce}}, \bibinfo {author}
  {\bibfnamefont {M.~P.}\ \bibnamefont {Lilly}}, \bibinfo {author}
  {\bibfnamefont {T.}~\bibnamefont {Karaulanov}}, \bibinfo {author}
  {\bibfnamefont {A.}~\bibnamefont {Jarmola}}, \bibinfo {author} {\bibfnamefont
  {A.}~\bibnamefont {Laraoui}},\ and\ \bibinfo {author} {\bibfnamefont {V.~M.}\
  \bibnamefont {Acosta}},\ }\bibfield  {title} {\bibinfo {title} {Time-resolved
  diamond magnetic microscopy of superparamagnetic iron-oxide nanoparticles},\
  }\href {http://arxiv.org/abs/2411.13087
  http://dx.doi.org/10.1021/acsnano.4c16703} {\bibfield  {journal} {\bibinfo
  {journal} {ACS Nano}\ }\textbf {\bibinfo {volume} {19}},\ \bibinfo {pages}
  {10048} (\bibinfo {year} {2025})}\BibitemShut {NoStop}%
\bibitem [{\citenamefont {Levine}\ \emph {et~al.}(2019)\citenamefont {Levine},
  \citenamefont {Turner}, \citenamefont {Kehayias}, \citenamefont {Hart},
  \citenamefont {Langellier}, \citenamefont {Trubko}, \citenamefont {Glenn},
  \citenamefont {Fu},\ and\ \citenamefont {Walsworth}}]{Levine2019}%
  \BibitemOpen
  \bibfield  {author} {\bibinfo {author} {\bibfnamefont {E.~V.}\ \bibnamefont
  {Levine}}, \bibinfo {author} {\bibfnamefont {M.~J.}\ \bibnamefont {Turner}},
  \bibinfo {author} {\bibfnamefont {P.}~\bibnamefont {Kehayias}}, \bibinfo
  {author} {\bibfnamefont {C.~A.}\ \bibnamefont {Hart}}, \bibinfo {author}
  {\bibfnamefont {N.}~\bibnamefont {Langellier}}, \bibinfo {author}
  {\bibfnamefont {R.}~\bibnamefont {Trubko}}, \bibinfo {author} {\bibfnamefont
  {D.~R.}\ \bibnamefont {Glenn}}, \bibinfo {author} {\bibfnamefont {R.~R.}\
  \bibnamefont {Fu}},\ and\ \bibinfo {author} {\bibfnamefont {R.~L.}\
  \bibnamefont {Walsworth}},\ }\bibfield  {title} {\bibinfo {title} {Principles
  and techniques of the quantum diamond microscope},\ }\href
  {https://www.degruyter.com/document/doi/10.1515/nanoph-2019-0209/html?lang=en}
  {\bibfield  {journal} {\bibinfo  {journal} {Nanophotonics}\ }\textbf
  {\bibinfo {volume} {8}},\ \bibinfo {pages} {1945} (\bibinfo {year}
  {2019})}\BibitemShut {NoStop}%
\bibitem [{\citenamefont {Barry}\ \emph {et~al.}(2020)\citenamefont {Barry},
  \citenamefont {Schloss}, \citenamefont {Bauch}, \citenamefont {Turner},
  \citenamefont {Hart}, \citenamefont {Pham},\ and\ \citenamefont
  {Walsworth}}]{Barry2020}%
  \BibitemOpen
  \bibfield  {author} {\bibinfo {author} {\bibfnamefont {J.~F.}\ \bibnamefont
  {Barry}}, \bibinfo {author} {\bibfnamefont {J.~M.}\ \bibnamefont {Schloss}},
  \bibinfo {author} {\bibfnamefont {E.}~\bibnamefont {Bauch}}, \bibinfo
  {author} {\bibfnamefont {M.~J.}\ \bibnamefont {Turner}}, \bibinfo {author}
  {\bibfnamefont {C.~A.}\ \bibnamefont {Hart}}, \bibinfo {author}
  {\bibfnamefont {L.~M.}\ \bibnamefont {Pham}},\ and\ \bibinfo {author}
  {\bibfnamefont {R.~L.}\ \bibnamefont {Walsworth}},\ }\bibfield  {title}
  {\bibinfo {title} {Sensitivity optimization for nv-diamond magnetometry},\
  }\href@noop {} {\bibfield  {journal} {\bibinfo  {journal} {Reviews of Modern
  Physics}\ }\textbf {\bibinfo {volume} {92}} (\bibinfo {year}
  {2020})}\BibitemShut {NoStop}%
\end{thebibliography}%

\end{document}